# The Design of Dual Band Stacked Metasurfaces Using Integral Equations

Jordan Budhu, *Member, IEEE*, Eric Michielssen, *Fellow, IEEE*, and Anthony Grbic, *Fellow, IEEE*

*Abstract*—An integral equation-based approach for the design of dual band stacked metasurfaces which are invariant in one-dimension is presented. The stacked metasurface will generate collimated beams at desired angles in each band upon reflection. The conductor-backed stacked metasurface consists of two metasurfaces (a patterned metallic cladding supported by a dielectric spacer) stacked one upon the other. The stacked metasurface is designed in three phases. First the patterned metallic cladding of each metasurface is homogenized and modeled as an inhomogeneous impedance sheet. An Electric Field Integral Equation (EFIE) is written to model the mutual coupling between the homogenized elements within each metasurface, and from metasurface to metasurface. The EFIE is transformed into matrix equations by the method of moments. The nonlinear matrix equations are solved at both bands iteratively resulting in dual band complex-valued impedance sheets. In the second phase, optimization is applied to transform these complex-valued impedance sheets into purely reactive sheets suitable for printed circuit board fabrication by introducing surface waves. In the third phase, the metallic claddings of each metasurface are patterned for full-wave simulation of the dual band stacked metasurface. Using this approach, two dual band stacked metasurfaces are designed.

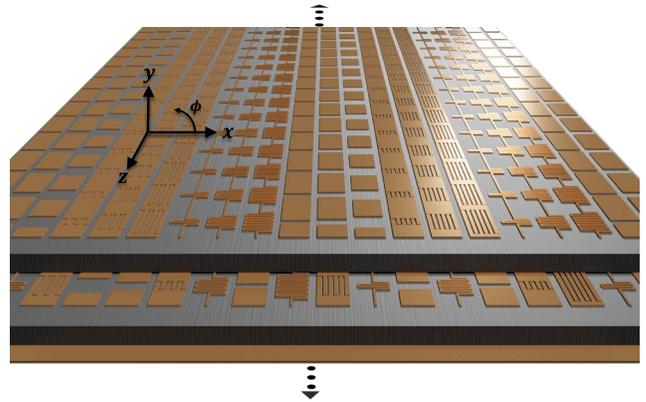

Figure 1. Dual band stacked metasurface. The geometry is infinite and invariant in the along the $z$-direction (wavenumber $k_z = 0$) and finite and spatially variant in the $x$- and $y$-directions. The electromagnetics problem is hence two-dimensional. The geometry contains a conductor backed stack of two metasurfaces (patterned metallic cladding supported by a dielectric spacer). The azimuthal angle $\phi$ lies in the $xy$-plane with reference axis $x$.

## I. INTRODUCTION

METASURFACES are the two-dimensional equivalents of metamaterials [1]. They are described by a boundary condition relating the average tangential electric and magnetic fields on the surface to the induced electric and magnetic surface currents flowing along it weighted by electric, magnetic, and magnetoelectric impedances or admittances. This boundary condition is called the Generalized Sheet Transition Condition (GSTC) [2,3]. Metasurfaces engineered using these boundary conditions can perform various specified electromagnetic transformations such as polarization rotation [4,5], circular polarization conversion [6], collimation [7], beamforming [8], and anomalous reflection [9-13]. In cases where dual band operation is desired, adding additional layers can result in operation at two different frequencies. When the surfaces are homogeneous, transfer matrix methods and network theory can be used to design the dual band metasurfaces with great accuracy. In these cases, the periodic metasurface is reduced to a single unit cell employing periodic boundaries. Network theory concepts are applied to the transmission/reflection through/from the single unit cell. For example, in [14], an analytic technique based on these methods was used to design dual band linear to circular polarizers for SatCom links. In [15-17], similar dual band circular polarizers are presented for operation in reflection. In [18], an optical dual band metasurface is designed for perfect optical antireflection at two Terahertz frequencies. In [19], machine learning based optimization is applied to design a unit cell geometry for homogeneous dual band frequency selective surfaces and wide band polarizing metasurfaces. All of these homogeneous metasurfaces are well modelled using periodic expansions and network theory. Homogeneous metasurfaces, however, only allow for a limited set of dual band field transformations typically involving polarization rotation, filtering, absorption, or antireflection.

Including inhomogeneity (non-periodicity) allows for more complex dual band field transformations such as refraction [20], collimation [21-24], focusing [25], and beamforming [26]. Modelling the mutual coupling between elements of these metasurfaces using local periodicity assumptions is

Jordan Budhu, Eric Michielssen, and Anthony Grbic are with the Electrical and Computer Engineering Department, University of Michigan, Ann Arbor, MI 48109 USA (e-mail: jbudhu@umich.edu, emichiel@umich.edu, agrbic@umich.edu).

This work was supported by the Army Research Office under Grant W911NF-19-1-0359, and the Office of Naval Research under Grant N00014-18-1-2536.

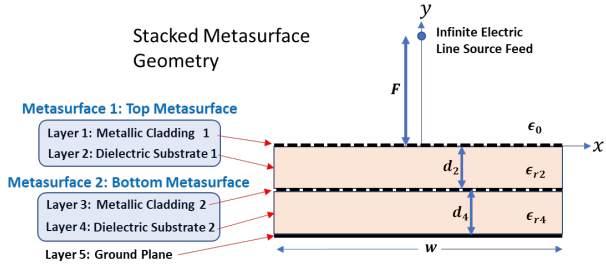

Figure 2. Dual band stacked metasurfaces geometry. The geometry is two-dimensional. The metasurface contains five layers: two metasurface layers (layers 1 and 3) and a ground plane (layer 5) separated by two dielectric spacers (layers 2 and 4). An infinite electric line source placed $F$ meters above the aperture feeds the metasurface.

approximate and the finite metasurface dimensions are not accounted for. The analytic network theory-based techniques applied to a unit cell geometry no longer provides accurate results since transverse coupling between cells is not modeled. Authors have proposed several methods to circumvent these problems. One such solution introduces perfect electric conducing baffles to separate the inhomogeneous surface into non-coupling homogeneous cells [27]. These surfaces contain many substrate vias which require advanced fabrication techniques and they cannot handle Transverse Magnetic polarization as perfect magnetic conductors are not available. Furthermore, if dual band metasurfaces are designed using local periodicity assumptions, the frequencies of operation must be chosen so that operation at one band does not interfere with that at the other band. This places unnecessary restrictions on the selection of operating frequencies. Authors have proposed several methods to circumvent this problem as well such as inserting a frequency selective surface between the layers containing elements operating at distinct bands [28]. This solution requires additional layers and does not model the mutual coupling accurately nor does it account for the finite metasurface dimensions.

Dual band inhomogeneous metasurfaces can be designed without resorting to local periodicity assumptions or network-based theories by modelling them with integral equations [29-33]. In [33], dual band metasurfaces are designed using integral equations and a least squares solution of the overdetermined system approach. However, the metasurfaces in [33] do not account for the truncated dielectric substrate and instead assume they are infinitely wide. The integral equation modeling approach presented in this paper accounts for all finite dimensions including the finite width and thickness of the metasurface and grounded substrates and also models correctly the mutual coupling between each of the unique elements in the inhomogeneous metasurface. We will describe the integral equation modelling approach and provide a scheme to solve these integral equations for dual band collimating metasurfaces. The collimated beams can be designed to scan to any far field angle. The presented scheme allows for great flexibility in the choice of operating frequencies and leads to the accurate prediction of the performance of dual band inhomogeneous metasurfaces that are invariant in one-dimension. The extension of the design approach to metasurfaces which vary in all three dimensions involves using the three-dimensional (3D) Green's function rather than the two-dimensional (2D) version and including the TM polarized excitations and associated induced currents in the moment method matrices.

In section II, a three phase design approach for the design of dual band metasurfaces is presented. In section III, the design approach is applied to the design of two dual band stacked metasurfaces. The first generates broadside beams at both 2.4GHz and 5.1GHz. The second generates a beam scanned to 30° off broadside when excited at 13.4GHz and a broadside beam when excited at 35.75GHz. The paper is concluded in section IV. An appendix provides numerical implementation details.

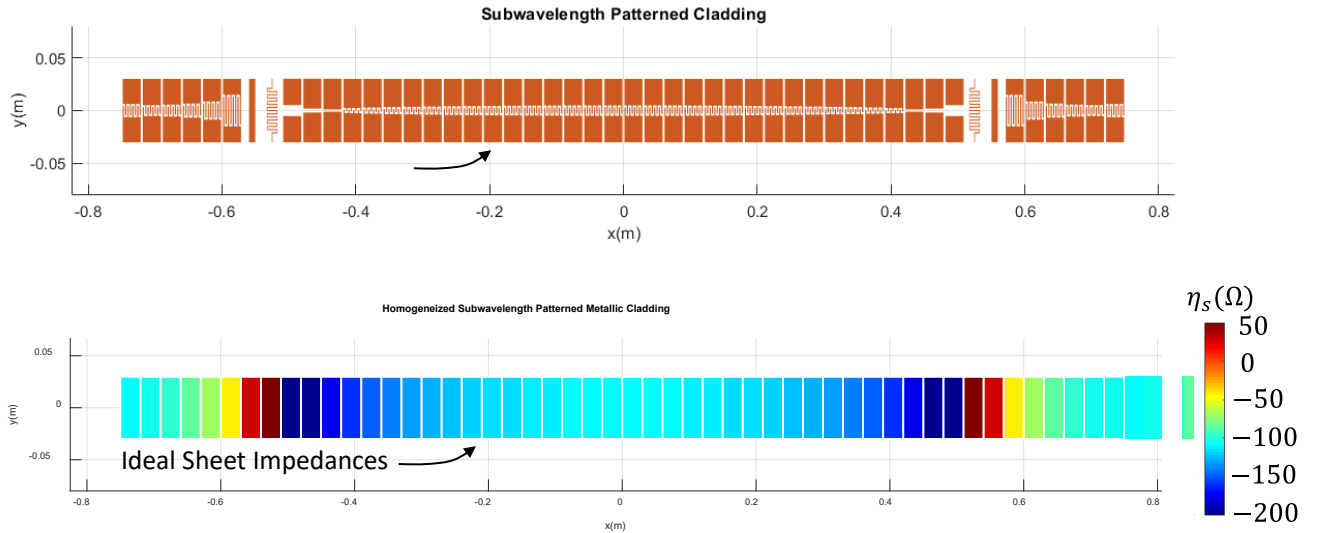

Figure 3. Homogenization of subwavelength patterned metallic cladding layers. The subwavelength patterned metallic copper elements are homogenized and represented by an ideal sheet impedance. The inhomogeneous impedance sheet made from the homogenized elements is used during design.

An $e^{j\omega t}$ time convention is used and suppressed throughout.

## II. DESIGN OF DUAL BAND STACKED METASURFACES

The dual band finite width stacked metasurface geometry considered consists of two metasurfaces stacked one atop the another (see Fig. 1). Each metasurface consists of a patterned metallic cladding layer supported by a dielectric spacer layer. The stack of metasurfaces is grounded by a finite width perfectly conducting backing layer. The stacked metasurface thus contains five total layers (see Fig. 2). The subwavelength texture of the patterned metallic claddings allows them to be homogenized. The patterned metallic claddings can therefore be modeled as inhomogeneous impedance sheets made of homogenized sheet impedance cells (see Fig. 3) according to the GSTC. In this case, the sheet impedances are assumed scalar given that the problem is 2D and only one polarization is considered. The dielectric spacers are modelled as volumetric distributions of electric polarization currents in free space following from the volume equivalence principle. A system of Electric Field Integral Equations (EFIE's) is written to account for mutual coupling between the unique homogenized cells within each layer (intra-layer) and from layer to layer (inter-layer). The system of EFIE's is transformed into a matrix equation by applying the method of moments discretization scheme.

The problem to solve is: given the incident and reflected fields, find the completely reactive sheet impedances necessary to obtain the desired field transformation. In the problem, both the sheet impedances and the induced surface currents on them are unknowns. As these quantities appear in the GSTC boundary condition as a product, the problem is inherently nonlinear. This is in contrast to the conventional forward problem of finding the reflected fields given the sheet impedances which is a linear problem. The solution to the problem is completed in three phases.

Phase 1 involves solving the non-linear matrix equations. A novel solution scheme has been developed to effectively linearize the matrix equations and solve them iteratively. The iterative scheme ping-pongs between frequencies at each iteration. Thus, the scheme designs the metasurface for operation at both frequencies simultaneously as it convergences (see Fig. 4 for preview). Consequently, there is great flexibility in the choice of each operating frequency used in design as elements operating at each band are not tuned independently but rather together. Phase 1 results in complex-valued sheet impedances (layers 1 and 3), since the specified transformations between the incident and reflected waves do not satisfy conservation of local power density at the metasurface planes [9-13]. Implementation of the real part of the complex-valued sheet impedances (with positive and negative resistances) adds cost and complexity to the fabrication given that active devices and attenuators are needed.

In phase 2, the real part is removed using an optimization approach originally introduced in [12]. The reactances (with resistances discarded) of the complex-valued sheet impedances (obtained in phase 1) are used as an initial point for gradient descent optimization. Since convergence of gradient descent algorithms strongly depends on obtaining a good initial guess, the results of phase 1 are paramount to the design process. The optimization cost function is formulated as the root-mean-square difference between the far field patterns radiated by the complex-valued sheet from phase 1 and that of the purely reactive sheet of the optimization. Hence, the reactive sheets will radiate the same far field amplitude pattern as the complex-valued sheets while conserving local power density at the sheets. This is achieved by introducing surface waves which redistribute the power transversally along the sheets. The purely reactive designs resulting can be fabricated using standard printed circuit techniques. Introducing surface waves to achieve passivity was introduced in [34].

In phase 3, the purely reactive sheet impedances of layers 1 and 3 obtained through optimization are realized as patterned metallic claddings. Full wave simulations of the final stacked metasurface show good agreement and validate the design process. At this stage, the problem is solved.

Although the specific implementation presented in this paper is two-dimensional, the idea behind the approach can also be applied to three-dimensional structures.

### A. Phase 1: Obtain Initial Metasurface Design Containing Complex-Valued Sheets

Consider a conductor backed stack of two metasurfaces shown in Fig. 2. The top metasurface will be denoted metasurface 1 and will support surface current $J_T$ along its surface (layer 1). The bottom metasurface will be denoted metasurface 2 and will support surface current $J_B$ along its surface (layer 3). The stacked metasurface has width $w$ along $x$ and is invariant along $z$. The top (layer 2) and bottom (layer 4) dielectric layers have relative permittivities $\epsilon_{r2}$ and $\epsilon_{r4}$, and thicknesses $d_2$ and $d_4$, respectively. In what follows, the top impedance sheet (layer 1) will be donated as $\eta_T$ and is assumed to reside in the $y = 0$ plane and the bottom impedance sheet (layer 3) will be denoted as $\eta_B$ and is assumed to reside at $y = -d_2$. Thus, the locations along the top (layer 1) and bottom (layer 3) impedance sheets are denoted by $\vec{\rho} = \vec{\rho}_T = x\hat{x}$ and $\vec{\rho} = \vec{\rho}_B = x\hat{x} - d_2\hat{y}$ with $-w/2 < x < w/2$. The stacked metasurface is illuminated by a $z$-directed line source with current $I_0$ Amps that resides at $\vec{\rho}_f = F\hat{y}$ meters above layer 1

$$E^{inc}(\vec{\rho}, \omega) = \frac{-I_0 \omega \eta_0}{4c} H_0^{(2)}\left(\frac{\omega}{c}|\vec{\rho} - \vec{\rho}_f|\right) \quad (1)$$

where $\omega$ defines the angular frequency of excitation, $c$ is the speed of light in free space, $\eta_0$ is the wave impedance of free space, and $H_0^{(2)}(\ )$ denotes the Hankel function of the section kind of order zero.

The proposed algorithm determines the impedances $\eta_T(\vec{\rho}_T, \omega)$ and $\eta_B(\vec{\rho}_B, \omega)$ of the top and bottom impedance sheets that produce reflected beams with prescribed scan angle $\phi_a^{sca}$ at frequency $\omega_a$ and $\phi_b^{sca}$ at frequency $\omega_b$. These beams are realized by the *desired* scattered aperture fields

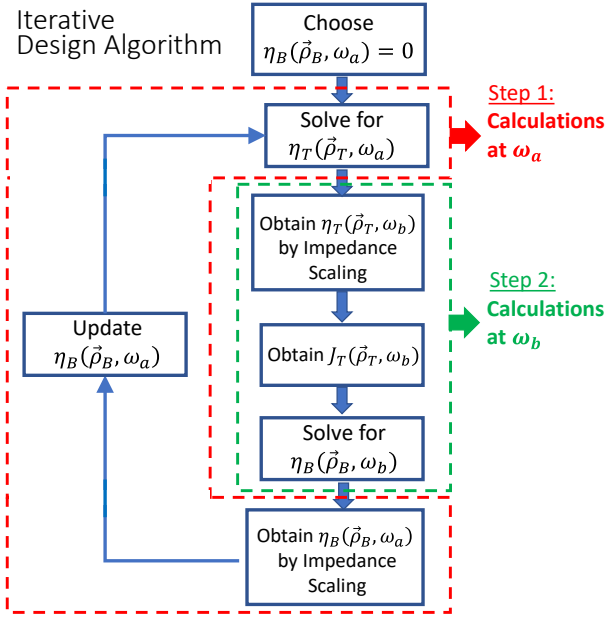

Figure 4. Iterative design scheme flowchart. The scheme involves solutions at two frequencies, $\omega_a$ and $\omega_b$. The calculations switch between the two frequencies during one pass of the iterative cycle. The quantities shown in red outline are calculations done at frequency $\omega_a$ while those in the green outline are calculations done at frequency $\omega_b$. At the transitions between frequencies, the impedances are scaled using the frequency dependence of the inductances and capacitances of the associated sheet impedance.

$$E_d^{sca}\left(\vec{\rho}_T, \omega\right) = \sqrt{\frac{2\eta_0 S^{inc}\left(\vec{\rho}_T, \omega\right)}{\sin\phi^{sca}}} e^{j\psi^{sca}(\omega)} \quad (2)$$

Where $S^{inc}$ is the incident field power density [12], $\psi^{sca}(\omega) = k_0 x \cos\phi^{sca}$, and $k_0$ is the wavenumber in free space. It follows from (1) and (2) that the total aperture field at the design frequencies is

$$E_d^{tot}\left(\vec{\rho}_T, \omega\right) = E^{inc}\left(\vec{\rho}_T, \omega\right) + E_d^{sca}\left(\vec{\rho}_T, \omega\right) \quad (3)$$

Let $J_T(\vec{\rho}_T, \omega)$ and $J_B(\vec{\rho}_B, \omega)$ denote electric current densities in the top and bottom impedance surfaces at frequency $\omega$. The total field in the bottom impedance sheet at $\omega$ is $E^{tot}(\vec{\rho}_B, \omega) = \eta_B(\vec{\rho}_B, \omega) J_B(\vec{\rho}_B, \omega)$. The current densities $J_T(\vec{\rho}_T, \omega)$ and $J_B(\vec{\rho}_B, \omega)$ therefore satisfy the coupled integral equations

$$E_d^{tot}\left(\vec{\rho}_T, \omega\right) = E\left(\vec{\rho}_T, J_T, \omega\right) + E\left(\vec{\rho}_T, J_B, \omega\right) + E^{inc}\left(\vec{\rho}_T, \omega\right)$$
$$\eta_B\left(\vec{\rho}_B, \omega\right) J_B\left(\vec{\rho}_B, \omega\right) = E\left(\vec{\rho}_B, J_T, \omega\right) + E\left(\vec{\rho}_B, J_B, \omega\right) + E^{inc}\left(\vec{\rho}_B, \omega\right)$$
$$(4)$$

where

$$E\left(\vec{\rho}, J, \omega\right) = -j\omega\mu_0 \int G\left(\vec{\rho}, \vec{\rho}', \omega\right) J\left(\vec{\rho}', \omega\right) d\vec{\rho}' \quad (5)$$

Here $G(\vec{\rho}, \vec{\rho}', \omega)$ is the Green's function for a current element radiating in the presence of the metasurfaces truncated (finite width) dielectric layers and ground plane, i.e.

$$\nabla^2 G\left(\vec{\rho}, \vec{\rho}', \omega\right) + k^2\left(\vec{\rho}, \omega\right) G\left(\vec{\rho}, \vec{\rho}', \omega\right) = -\delta\left(\vec{\rho}'\right) \quad (6)$$

where $k^2(\vec{\rho}, \omega) = \omega^2 \mu_0 \epsilon_0 \epsilon_r(\vec{\rho}, \omega)$. While no closed-form expression for $G(\vec{\rho}, \vec{\rho}', \omega)$ exists, it can be computed numerically as illustrated in the appendix.

To determine $\eta_T(\vec{\rho}_T, \omega)$ and $\eta_B(\vec{\rho}_B, \omega)$, the proposed algorithm initializes $\eta_B(\vec{\rho}_B, \omega) = 0$ and iteratively executes Steps 1 and 2 below until convergence occurs. This iterative scheme is depicted in Fig. 4. The numerical implementation details of the algorithm can be found in the appendix. Note, convergence is achieved when the scattered far field patterns resulting from the metasurface at both frequencies does not change with subsequent iterations

**Step 1: Compute $\eta_T(\vec{\rho}_T, \omega)$ by enforcing aperture condition (3) at $\omega_a$ given $\eta_B(\vec{\rho}_B, \omega)$.**

In terms of the numerical Green's functions $[G_{TT}], [G_{TB}], [G_{BT}]$, and $[G_{BB}]$ derived in the appendix, (4) evaluated at $\omega = \omega_a$ is expressed as

$$[W_T^{\omega_a}] = [G_{TT}][I_T^{\omega_a}] + [G_{TB}][I_B^{\omega_a}] + [V_T^{\omega_a}]$$
$$[\eta_B^{\omega_a}][I_B^{\omega_a}] = [G_{BT}][I_T^{\omega_a}] + [G_{BB}][I_B^{\omega_a}] + [V_B^{\omega_a}] \quad (7)$$

The matrix definitions in (7) can be found in the appendix. First, solve the system of equations (two equations two unknowns) (7) for the two unknown currents $[I_T^{\omega_a}]$ and $[I_B^{\omega_a}]$. Note, $[I_T^{\omega_a}]$ and $[I_B^{\omega_a}]$ are the complex coefficients vector of the expansions of $J_T(\vec{\rho}_T, \omega_a)$ and $J_B(\vec{\rho}_B, \omega_a)$ as in (A.3). Then obtain the sheet impedances of the top metasurfaces from

$$\left[\eta_T^{\omega_a}\right] = \frac{[W_T^{\omega_a}]}{[I_T^{\omega_a}]} \quad (8)$$

It should be noted that the practical realization of $\eta_T(\vec{\rho}_T, \omega_a)$ fixes its value at all other frequencies, including $\omega_b$. The full wave simulation results for the frequency scaling of the patterned elements used in phase 3 show that they follow the usual frequency dependance relations for lumped inductive and capacitive reactances (see Fig. 6). Thus, next frequency-scale the top metasurfaces impedance from $\omega_a$ to $\omega_b$. The capacitive reactances are scaled up as

$$X^{\omega_b} = \frac{\omega_a}{\omega_b} X^{\omega_a} \quad (9)$$

and the inductive reactances are scaled down as

$$X^{\omega_b} = \frac{\omega_b}{\omega_a} X^{\omega_a} \quad (10)$$

**Step 2: Compute $\eta_B(\vec{\rho}_B, \omega)$ by enforcing aperture condition (3) at $\omega_b$ given $\eta_T(\vec{\rho}_T, \omega)$ from step 1.**

Assuming that $\eta_T(\vec{\rho}_T, \omega_b)$ is known, we may compute $J_T(\vec{\rho}_T, \omega_b)$ from knowledge of the total aperture field (3) at $\omega_b$ using

$$\left[I_T^{\omega_b}\right] = \frac{[W_T^{\omega_b}]}{[\eta_T^{\omega_b}]} \quad (11)$$

Once $J_T(\vec{\rho}_T, \omega_b)$ is known, an integral equation for $J_B(\vec{\rho}_B, \omega_b)$ is constructed by expressing the known total aperture field at $\omega_b$ as the sum of fields produced by both currents.

$$[W_T^{\omega_b}] = [G_{TT}][I_T^{\omega_b}] + [G_{TB}][I_B^{\omega_b}] + [V_T^{\omega_b}] \quad (12)$$

After solving this equation for $J_B(\vec{\rho}_B, \omega_b)$, the total field on the impedance surface of the bottom metasurface (layer 3), $E^{tot}(\vec{\rho}_B, \omega_b)$, is computed as

$$\left[W_B^{\omega_b}\right] = [G_{BT}]\left[I_T^{\omega_b}\right] + [G_{BB}]\left[I_B^{\omega_b}\right] + \left[V_B^{\omega_b}\right] \quad (13)$$

Knowledge of both $E^{tot}(\vec{\rho}_B, \omega_b)$ and $J_B(\vec{\rho}_B, \omega_b)$ finally allows $\eta_B(\vec{\rho}_B, \omega_b)$ to be computed via

$$\left[\eta_B^{\omega_b}\right] = \frac{\left[W_B^{\omega_b}\right]}{\left[I_B^{\omega_b}\right]} \quad (14)$$

The practical realization of $\eta_B(\vec{\rho}_B, \omega_b)$ fixes its value at all other frequencies, including $\omega_a$. Thus, next frequency scale $\eta_B(\vec{\rho}_B, \omega_b)$ to $\eta_B(\vec{\rho}_B, \omega_a)$. As before, the impedances follow the usual scaling laws for inductive and capacitve reactances. Thus, the capacitive reactances are scaled down as

$$X^{\omega_a} = \frac{\omega_b}{\omega_a} X^{\omega_b} \quad (15)$$

and the inductive reactances are scaled up as

$$X^{\omega_a} = \frac{\omega_a}{\omega_b} X^{\omega_b} \quad (16)$$

Finally, update $\eta_B(\vec{\rho}_B, \omega_a)$ by interpolating the new value with the previous as $\eta_B(\vec{\rho}_B, \omega_a) = s_1 \eta_B(\vec{\rho}_B, \omega_a)_{new} + s_2 \eta_B(\vec{\rho}_B, \omega_a)_{previous}$, where $s_1$ and $s_2$ are interpolation constants ($s_1 = 0.2$ and $s_2 = 0.8$).

These 2 steps are repeated until convergence is achieved.

### B. Phase 2: Optimization of Initial Complex-Valued Sheets to Obtain Purely Reactive Sheets

The real part of the converged complex-valued sheet impedances of (8) and (14) indicate the need for lossy (positive resistances) or active (negative resistances) elements. Implementation of lossy/active devices is costly and difficult to implement with standard printed circuit board fabrication techniques. A purely reactive metasurface is desired to simplify the realization. An optimization method based on gradient descent is next used to optimize the impedance sheets and obtain purely reactive sheet impedances. The optimization scheme is depicted in Fig. 5. Given that the patterns produced by the stacked metasurface containing the complex-valued sheet impedances obtained in phase 1 represent the desired performance, the cost function for the optimization is

$$\begin{aligned} f_{cost} = &\ RMS\left\{\left|E_{farfield}^{\omega_a}(\phi)\right|_{\substack{Complex\\Sheet}} - \left|E_{farfield}^{\omega_a}(\phi)\right|_{\substack{Reactive\\Sheet}}\right\} + \\ &\ RMS\left\{\left|E_{farfield}^{\omega_b}(\phi)\right|_{\substack{Complex\\Sheet}} - \left|E_{farfield}^{\omega_b}(\phi)\right|_{\substack{Reactive\\Sheet}}\right\} \end{aligned} \quad (17)$$

where $\left|E_{farfield}^{\omega_a}(\phi)\right|$ and $\left|E_{farfield}^{\omega_b}(\phi)\right|$ is the magnitude of the scattered far field pattern at frequency $\omega_a$ and $\omega_b$, respectively. Note, RMS is the root-mean-square value of the quantity in the brackets. The cost function attempts to minimize the difference between the far field pattern radiated by the reactive sheet and the complex sheet. The optimization variables are the reactances of the elements themselves. Thus, the optimizer is seeded by throwing out the real part of the complex sheet impedances and keeping only the reactances.

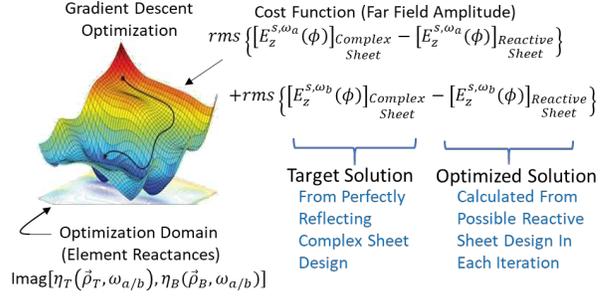

Figure 5. Gradient descent optimization. The unknown reactances are arranged along orthogonal axes to create the optimization domain. The cost function is defined in this space as a function of these unknown reactances as the root-mean-square difference between the far fields radiated by the metasurface containing the complex-valued sheet impedances and that of the metasurface containing the purely reactive sheet impedances (optimization variables).

The optimization is constrained to reactances which are practically achievable through the patterning of subwavelength cells of metallic claddings. From Fig. 6, these limits are $+j150\Omega$ for inductive reactances and $-j5000\Omega$ for capacitive reactances. The optimization continues until (17) is minimized.

### C. Phase 3: Realization of Purely Reactive Sheets Through Patterning of the Metallic Claddings

In phase 3, the optimized reactive sheets are realized through patterning of the metallic claddings. Five element topologies were utilized. They are shown in Figs. 5a and 5b and include: a Gap Capacitor (Gap Cap), an Interdigitated Capacitor (IDC), an element made with Parallel Shunt LC Resonators (Para Res), a Straight-Line Inductor (Str Ind), and a Meandered Line Inductor (Meand). The Para Res element was designed to short at both frequencies simultaneously so the elements follow the frequency scaling of (9), (10), (15), and (16) used in design. The element unit cells are 0.0029m ($\lambda_b/20$) wide in the $x$-direction by 0.0059m ($\lambda_b/10$) tall in the $z$-direction. The metallization within the element cell is a maximum of 0.0026m ($0.9 \times \lambda_b/20$) wide in order to electrically isolate them from their nearest neighbors. There are two media to consider. The elements on layer 1 sit at an interface between a half space of air and a half space of dielectric. This case is shown in the inset of Fig. 6a. On the other hand, the elements on layer 3 sit within a dielectric medium as shown in the inset of Fig. 6b. Extractions are run separately for each case. The element impedances at both frequencies are shown in Fig. 6 for both the elements on layer 1 (Fig. 6a) and the elements on layer 3 (Fig. 6b). Although the design thus far (phase 1 and 2) was done without assuming any local periodicity, these impedances were extracted in a locally periodic environment by calculating the $Z_{12}$ element of the impedance matrix. This produces a small error which will be corrected through direct optimization of the cladding in full-wave simulation (see Results section). Also plotted in Fig. 6a and 6b are the ideal scaling laws for inductive and capacitive reactances of (9), (10), (15), and (16) shown with solid black lines. The elements follow the assumed scaling laws closely.

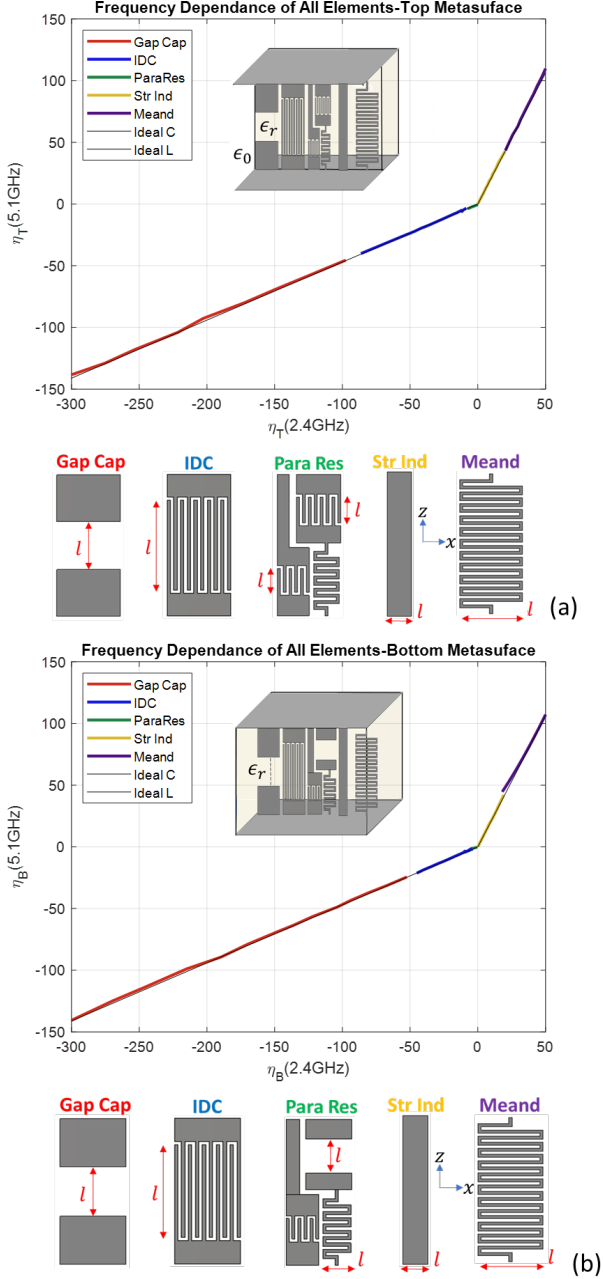

Figure 6. Element geometry extraction. Five types of elements are used. A Gap Capacitor (Gap Cap), an Interdigitated Capacitor (IDC), a parallel shunt LC resonator (Para Res), a Straight-Line Inductor (Str Ind), and a Meandered Line Inductor (Meand). Shown are the extraction curves for both the top metasurface (a) (where the elements sit on the interface of an infinite half space of air and an infinite half space of dielectric with permittivity $\epsilon_r$) and the bottom metasurface (b) (where the elements sit within an infinite medium of permittivity $\epsilon_r$). The frequency scaling for the elements are shown in the figures and approximately follow the usual scaling laws for reactances.

## III. RESULTS

*Design Example 1: Dual band Design for Wi-Fi Bands*

In this design, two popular Wi-Fi bands are serviced by the same antenna. The design operates at $f_a$=2.4GHz and $f_b$=5.1GHz. The metasurface has dimensions $w$=0.2529m. The line source was placed at a focal length of $F$=0.25m giving an F/D ratio of approximately 1. The dielectric spacers are Rogers 6010 substrates with $\epsilon_{r1} = \epsilon_{r2} = 10.7(1 - j0.0023)$. The dielectric spacer thicknesses are $d_2$=$d_4$=1.27mm. The integral equations given by (3) are solved by the method of moments (see Appendix A for details). $N_1$=$N_2$=$N_3$=86 basis and expansion functions were used in the moment method algorithm for the surfaces and $N_2$=$N_4$=1500 basis and expansion functions for the dielectric volumes. A small gap was introduced between the elements to electrically isolate them by setting the surface current basis function width to 0.9 times the spacing between the basis centers.

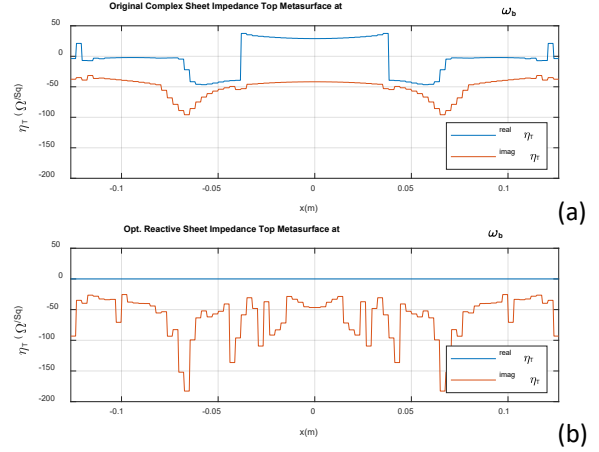

Figure 7. Sheet impedance for top metasurface at $\omega_b$. (a) The complex sheet impedance resulting from the direct application of the algorithm in section II.A. (b) The purely reactive sheet impedance resulting from the optimization algorithm in section II.B.

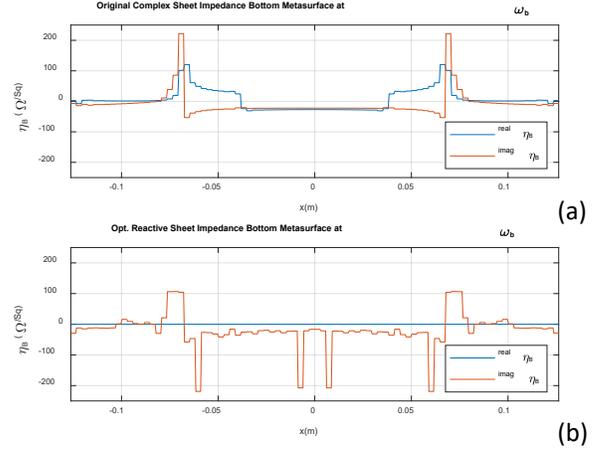

Figure 8. Sheet impedance for bottom metasurface at $\omega_b$. (a) The complex sheet impedance resulting from the direct application of the algorithm in section II.A. (b) The purely reactive sheet impedance resulting from the optimization algorithm in section II.B.

The converged surface impedances computed in phase 1 for layer 1 and layer 3 are shown in Figs. 7a and 8a. The surface impedances at $\omega_a$ can be obtained through the assumed frequency scaling laws of (9), (10), (15), and (16) used during design. Convergence was achieved in less than 100 iterations. The far field patterns resulting from the stacked metasurface containing the complex-valued sheets calculated using the method of moments code at each frequency are shown in Fig. 9 and Fig. 10 as the curves labeled 'MoM Complex'. The imaginary part of Fig. 7a and Fig. 8a serve as the initial guess used in the gradient descent optimizer in phase 2. The far field patterns labeled 'MoM Complex' in Fig. 9 and Fig. 10 serve as the $\left|E_{farfield}^{\omega_a}(\phi)\right|_{complex\ sheet}$ and $\left|E_{farfield}^{\omega_b}(\phi)\right|_{complex\ sheet}$ in (17).

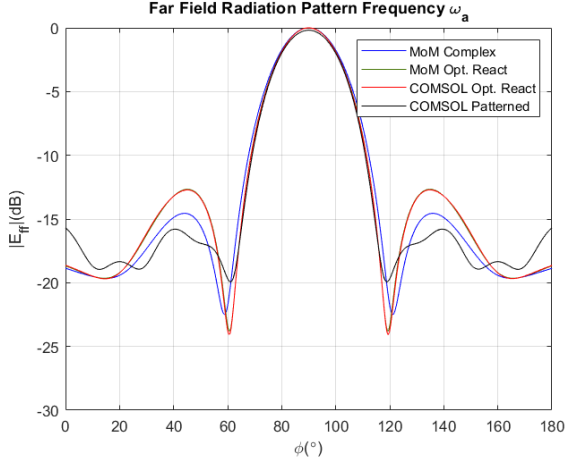

Figure 9. Far field pattern at frequency $\omega_a$. The far field pattern resulting from the metasurface made from the homogenized complex valued impedance sheets is shown in the blue curve (MoM Complex). The pattern resulting from the optimized homogenized purely reactive impedance sheet is shown in the green (MoM Opt. React) and red (Comsol Opt. React) curves. The far field pattern resulting from the metasurface made from patterned metallic cladding layers is shown in the black curve (Comsol Patterned).

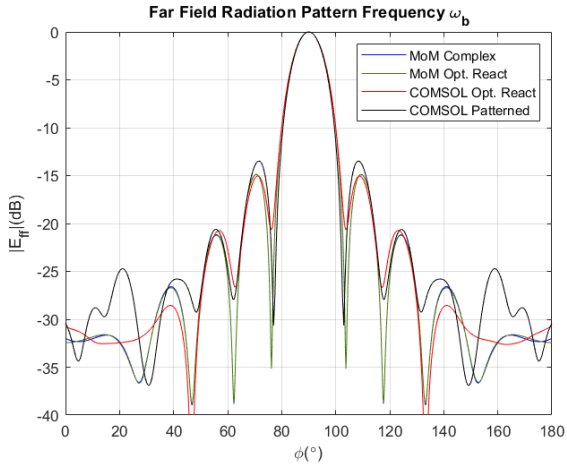

Figure 10. Far field pattern at frequency $\omega_b$. The far field pattern resulting from the metasurface made from the homogenized complex valued impedance sheets is shown in the blue curve (MoM Complex). The pattern resulting from the optimized homogenized purely reactive impedance sheet is shown in the green (MoM Opt. React) and red (Comsol Opt. React) curves. The far field pattern resulting from the metasurface made from patterned metallic cladding layers is shown in the black curve (Comsol Patterned).

The optimized purely reactive sheet impedances of phase 2 are shown in Fig. 7b and Fig. 8b. It should be noted that no inductive reactances exceed $+j150\Omega$ and no capacitive reactances exceed $-j5000\Omega$, as these are the limits imposed as constraints during the optimization.

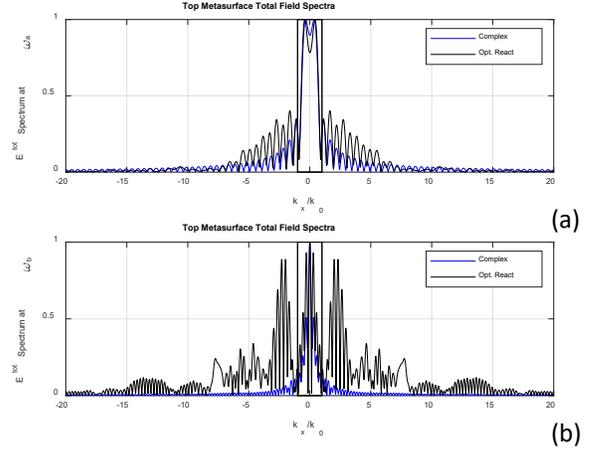

Figure 11. The amplitude of the total field spectrum resulting from the direct application of the algorithm in section II.A in the blue curve (Complex) and from the optimization algorithm in section II.B in the red curve (Opt. React) for the top metasurface. (a) For $\omega_a$ and (b) for $\omega_b$.

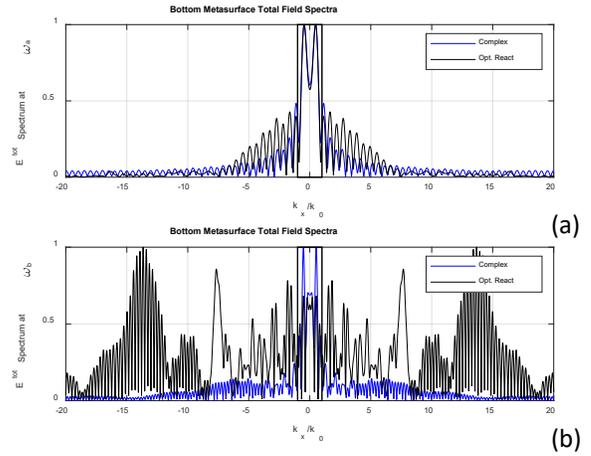

Figure 12. The amplitude of the total field spectrum resulting from the direct application of the algorithm in section II.A in the blue curve (Complex) and from the optimization algorithm in section II.B in the red curve (Opt. React) for the bottom metasurface. (a) For $\omega_a$ and (b) for $\omega_b$.

In Figs. 11 and 12, the plane wave spectrum of the total electric field along the top metasurface (at $\omega_a$ and $\omega_b$) and along the bottom metasurface (at $\omega_a$ and $\omega_b$) is obtained from

$$\widetilde{E}_i^{tot}(k_x, \omega_{a/b}) = \Im\left[E_i^{tot}(\vec{\rho}_i, \omega_{a/b})\right] \\ = \Im\left[\eta_i(\vec{\rho}_i, \omega_{a/b}) J_i(\vec{\rho}_i, \omega_{a/b})\right] \quad (18)$$

In (18), $\Im$ is the Fourier transform operator. Note, $i$ indicates which metasurface can take values $i = T$ or $B$. The plane-wave spectrum is plotted for the cases of the complex sheet and the optimized reactive sheet. As can be seen, the optimization technique has introduced significant evanescent spectrum. The evanescent spectrum is associated with surface waves that redistribute power along the metasurface planes in order to achieve passivity [12,34].

In Figs. 9 and 10, the far field patterns generated by the optimized reactive sheets are shown superimposed over those generated by the complex sheets. The patterns were calculated using both MoM code and the commercial finite element electromagnetics solver COMSOL Multiphysics as a full-wave verification. The agreement is excellent.

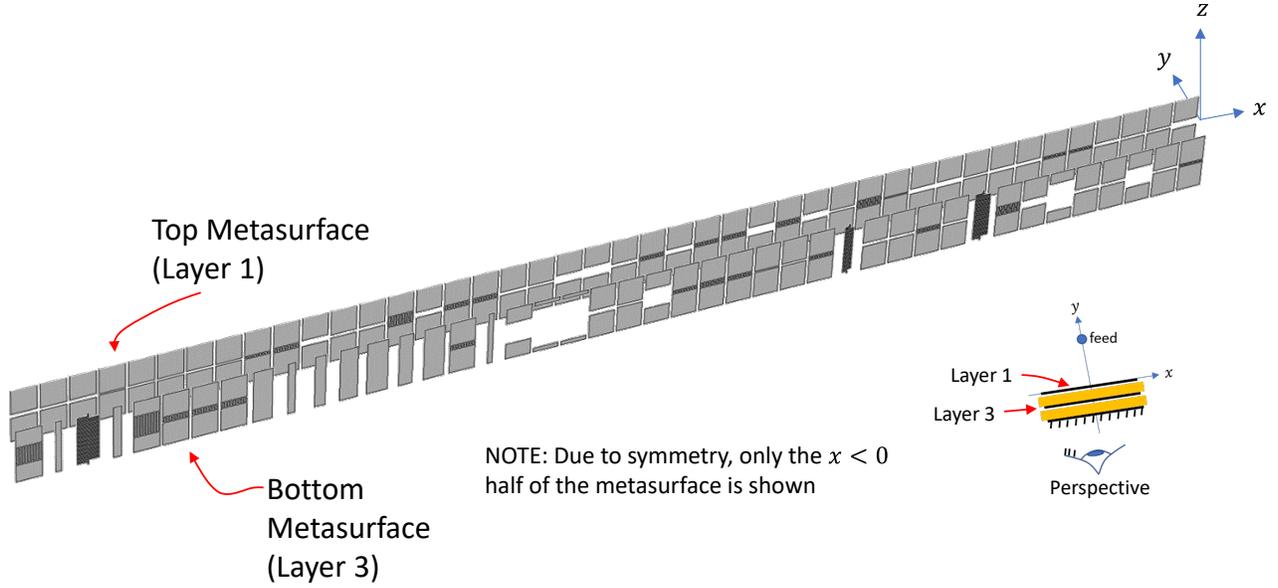

Figure 13. Stacked metasurface patterned metallic claddings. Only layers 1 and 3 are shown (the dielectric spacers and ground plane are not shown). The perspective is shown in the inset. The metasurface is being viewed from behind layer 3 with layer 3 the first layer visible. These patterned layers are placed within a parallel plate waveguide in order to create an effective 2-dimensional structure infinite in the $z$-direction.

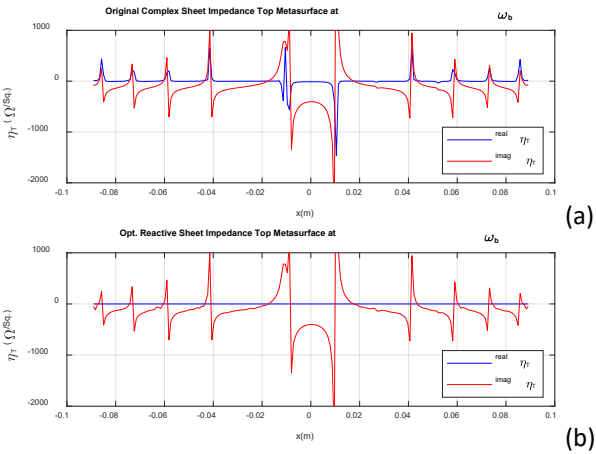

Figure 14. Sheet impedance for top metasurface at $\omega_b$. (a) The complex sheet impedance resulting from the direct application of the algorithm in section II.A. (b) The purely reactive sheet impedance resulting from the optimization algorithm in section II.B.

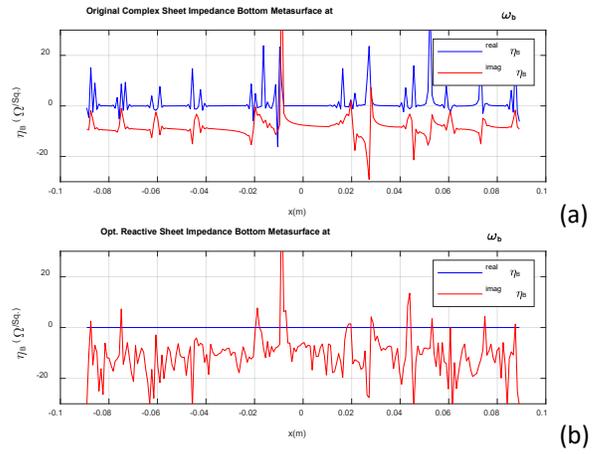

Figure 15. Sheet impedance for bottom metasurface at $\omega_b$. (a) The complex sheet impedance resulting from the direct application of the algorithm in section II.A. (b) The purely reactive sheet impedance resulting from the optimization algorithm in section II.B.

In phase 3, the metallic claddings of both metasurfaces are patterned according to the impedances defined in Fig. 7b and Fig. 8b. Because these impedances vary in a non-adiabatic way and because the extraction of the impedances associated with the patterned geometry was done in a periodic environment, the extracted $Z_{12}$ will be approximate and will require subsequent fine tuning optimization of the patterned metallic cladding. The geometric parameter $l$ for each of the elements of the patterned metallic claddings (see Fig. 6a and Fig. 6b) were optimized using gradient descent optimization with the same cost function (17) except the reactive sheets are replaced by the patterned claddings. The initial patterned metallic claddings built from Fig. 7b and Fig. 8b provide a very good initial guess and hence the optimization converges rapidly. The final patterned metallic claddings are shown in Fig. 13. In the figure, layer 3 is visible first with layer 1 shown behind it. The patterned stacked metasurface was placed within a Parallel Plate Waveguide (PPW) and simulated in COMSOL Multiphysics at both frequencies. The resulting far field patterns are plotted superimposed over the method of moments results of the homogenized sheets version in Fig. 9 and Fig. 10. As the figure shows, the patterned metasurfaces show excellent agreement with the theoretical results after the fine tuning of the claddings thus validating the design technique, element characterization,

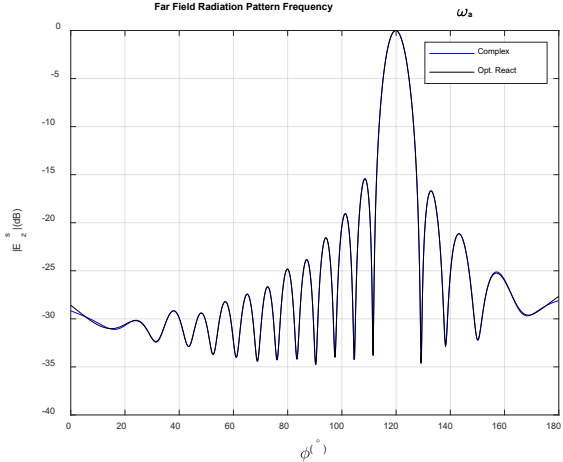

Figure 16. Far field pattern at frequency $\omega_a$. The far field pattern resulting from the original complex sheet is shown in the blue curve (Complex). The far field pattern resulting from the optimized purely reactive sheet is shown in the black curve (Opt. React).

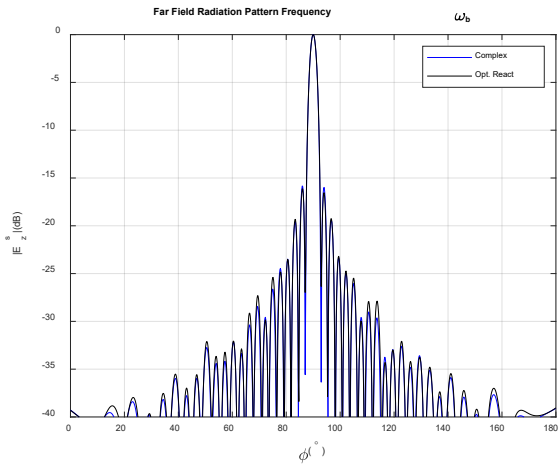

Figure 17. Far field pattern at frequency $\omega_b$. The far field pattern resulting from the original complex sheet is shown in the blue curve (Complex). The far field pattern resulting from the optimized purely reactive sheet is shown in the black curve (Opt. React).

and patterning process.

*Design Example 2: Dual band Design for Ka-Band and Ku-Band*

This design operates at both $f_a$=13.4GHz and $f_b$=35.75GHz. The lower band beam is scanned to 30° off broadside ($\psi^{sca}(x) = k_0 x cos60°$ in (2)) while the upper band beam is directed broadside ($\psi^{sca}(x) = k_0 x cos90°$ in (2)). The metasurface has dimensions $w$=0.1791m. The line source was placed at a focal length of $F$=0.0896m giving an F/D ratio of 0.5. The dielectric spacers are Rogers RT/Duroid 5880 substrates with $\epsilon_{r2} = \epsilon_{r4} = 2.2(1 - j0.0009)$. The dielectric spacer thicknesses are $d_2$=0.56mm and $d_4$=0.224mm. The integral equations given by (3) were solved by the method of moments (see Appendix A for details). $N_1$=$N_2$=$N_3$=214 basis and expansion functions were used in the moment method algorithm for the surfaces and $N_2$=$N_4$=1070 basis and expansion functions for the dielectric volumes.

The converged surface impedances computed in phase 1 for layer 1 and layer 3 at $\omega_b$ are shown in Fig. 14a and Fig. 15a. Convergence was achieved in less than 10 iterations. The far field patterns calculated with the method of moments code for each frequency are shown in Fig. 16 and Fig. 17 as the curves labeled 'Complex'.

The optimized purely reactive sheet impedances from phase 2 are shown in Fig. 14b and Fig. 15b. The far field patterns resulting from the optimized reactive sheets are shown superimposed in Fig. 16 and Fig. 17. The agreement is excellent. Since a patterned design was already shown, this second design will not be realized as a patterned metasurface. This design shows the versatility of the presented design algorithm to obtain scanned collimated beams at arbitrarily defined frequencies.

## IV. Conclusion

A technique to design dual band stacked metasurfaces was presented. The conductor backed stacked metasurface considered consists of two metasurfaces (a patterned metallic cladding supported by a dielectric spacer) stacked one upon the other. The stacked metasurface is modeled using coupled volume surface integral equations. These integral equations are transformed into matrix equations by the method of moments. The matrix equations are nonlinear as the product of induced current density and unknown sheet impedance for the bottom metasurface cannot be related to a stipulated total field. An iterative design algorithm was developed to solve the nonlinear system of integral equations. The iterative design scheme allows for flexibility in the specifiaction of the two operating frequencies. The stacked metasurfaces are designed in three phases. Phase 1 obtains a complex-valued impedance sheet design. In phase 2, the complex-valued sheet impedaces are transformed into purely reactive sheet impedances through optimization. In phase 3, the purely reactive sheets are realized as patterned metallic claddings. Two design examples were reported. The first was a 2.4GHz/5.1GHz dual band stacked metasurface. The stacked metasurface was patterned and simulated using COMSOL Multiphysics full-wave simulation tools at both bands. The results agree well with the metasurface made from homogenized sheets. The second is a Ku-band/Ka-band dual band stacked metasurface. The far field patterns of the optimized reactive design agree well with the original complex sheet design.

## Appendix: Method of Moment Solution Details

### A. Construction of Electric Field Integral Equations and Method of Moment Matrix Equations

Expanding (3) for the first layer ($i$=1), the EFIE becomes

$$E^{inc}(x,0) - j\frac{\eta_0 \omega}{c} \int_{-w/2}^{w/2} G(x,0;x',0) J_1(x',0) dx'$$

$$- j\frac{\eta_0 \omega}{c} \int_{-d_2/2}^{d_2/2} \int_{-w/2}^{w/2} G(x,0;x',y') J_2(x',y') dx' dy'$$

$$- j\frac{\eta_0 \omega}{c} \int_{-w/2}^{w/2} G(x,0;x',-d_2) J_3(x',-d_2) dx'$$

$$- j\frac{\eta_0 \omega}{c} \int_{-d_4/2}^{d_4/2} \int_{-w/2}^{w/2} G(x,0;x',y') J_4(x',y') dx' dy'$$

$$- j\frac{\eta_0 \omega}{c} \int_{-w/2}^{w/2} G(x,0;x',-d_2-d_4) J_5(x',-d_2-d_4) dx'$$

$$= \eta_1(x,0) J_1(x,0)$$

(A.1)

where $J_i$ is the electric surface current density on layer $i$ when $i=1,3,5$, or the electric volume current density within layer $i$ when $i=2,4$. The 2-dimensional Green's Function is given by

$$G(x,y;x',y') = \frac{1}{4j} H_0^{(2)}\left(\frac{\omega}{c}\sqrt{(x-x')^2 + (y-y')^2}\right) \quad (A.2)$$

Note, the primed coordinates represent the source points and the unprimed the observation coordinates. There are two types of expansion functions for the two types of currents. For the surface currents, 1-dimensional (1D) pulse basis functions are used. For the volumetric currents, 2-dimensional (2D) pulse basis functions are used. The surface currents on layer $i$ ($i=1,3,5$) are expanded into $N_i$ 1D pulse basis functions each of width $\Delta_x = w_i/N_i$

$$J_i(x') = \sum_{n=0}^{N_i-1} I_n P_n(x') \quad (A.3)$$

$$P_n(x') = \begin{cases} 1, & \text{if } |x-x'| < \Delta_x/2 \\ 0, & \text{otherwise} \end{cases} \quad (A.4)$$

The volume currents on layer $i$ ($i=2,4$) are expanded into $N_i$ 2D pulse basis functions each of area $\Delta_x \times \Delta_y = w_i d_i/N_i$

$$J_i(x',y') = \sum_{n=0}^{N_i-1} I_n P_n(x',y') \quad (A.5)$$

$$P_n(x',y') = \begin{cases} 1, & \text{if } |x-x'| < \Delta_x/2,\ |y-y'| < \Delta_y/2 \\ 0, & \text{otherwise} \end{cases} \quad (A.6)$$

Substituting (A.2), (A.3), and (A.5) into (A.1) and using a Galerkin testing procedure, the integral equation in (A.1) can be discretized and written in matrix form as

$$[V_1] - [Z_{11}][I_1] - [Z_{12}][I_2] - [Z_{13}][I_3] \\ - [Z_{14}][I_4] - [Z_{15}][I_5] = [\eta_1][I_1] \quad (A.7)$$

where the matrices $[Z_{ij}]$ are given by (A.8). Here, $(x_m, y_m)$ and $(x_n, y_n)$ denote the coordinates of the centroid of testing function $m$ and basis function $n$, respectivey. The matrix element $[Z_{ij}(m,n)]$ representes the mutual impedance between basis function $n$ in layer $j$ and test function $m$ in layer $i$. The self-impedance term occurs whenever $i=j$ and $m=n$ simultaneously in (A.8). In this case, the impedance matrix element is found from

$$Z_{ii}(n,n) = \frac{\eta_0 \omega \Delta_x^2}{4c} + j\frac{\eta_0 \omega}{4\pi c}\Delta_x^2\left[3 + \ln(4) - 2\ln\left(\frac{\gamma\omega\Delta_x}{c}\right)\right] (A.9)$$

where $\gamma=1.781$ and is Euler's number. The integral equations for the third and fifth layer are similar to (A.1) as they are EFIE's constructed over a surface. For the second and fourth layer, the EFIE's are constructed throughout a volume. For example, the integral equation for layer 2 can be written as

$$E_z^i(x,y) - j\frac{\eta_0 \omega}{c}\int_{-w/2}^{w/2} G(x,y;x',0) J_1(x',0) dx'$$

$$- j\frac{\eta_0 \omega}{c}\int_{-d_2/2}^{d_2/2}\int_{-w/2}^{w/2} G(x,y;x',y') J_2(x',y') dx' dy'$$

$$- j\frac{\eta_0 \omega}{c}\int_{-w/2}^{w/2} G(x,y;x',-d_2) J_3(x',-d_2) dx'$$

$$- j\frac{\eta_0 \omega}{c}\int_{-d_4/2}^{d_4/2}\int_{-w/2}^{w/2} G(x,y;x',y') J_4(x',y') dx' dy'$$

$$- j\frac{\eta_0 \omega}{c}\int_{-w/2}^{w/2} G(x,y;x',-d_2-d_4) J_5(x',-d_2-d_4) dx'$$

$$= \eta_2(x,y) J_2(x,y)$$

(A.10)

where $-d_2 < y < 0$. Substituion of (A.2), (A.3), and (A.5) into

$$[Z_{11}] = \frac{\eta_0 \omega}{4c} \int_{x_{m1}-\Delta_x/2}^{x_{m1}+\Delta_x/2} \int_{x_{n1}-\Delta_x/2}^{x_{n1}+\Delta_x/2} H_0^{(2)}\left(\frac{\omega}{c}\sqrt{(x-x')^2}\right) dx' dx$$

$$[Z_{12}] = \frac{\eta_0 \omega}{4c} \int_{x_{m1}-\Delta_x/2}^{x_{m1}+\Delta_x/2} \int_{x_{n2}-\Delta_x/2}^{x_{n2}+\Delta_x/2} \int_{y_{n2}-\Delta_y/2}^{y_{n2}+\Delta_y/2} H_0^{(2)}\left(\frac{\omega}{c}\sqrt{(x-x')^2 + (0-y')^2}\right) dy' dx' dx$$

$$[Z_{13}] = \frac{\eta_0 \omega}{4c} \int_{x_{m1}-\Delta_x/2}^{x_{m1}+\Delta_x/2} \int_{x_{n3}-\Delta_x/2}^{x_{n3}+\Delta_x/2} H_0^{(2)}\left(\frac{\omega}{c}\sqrt{(x-x')^2 + d_2^2}\right) dx' dx \quad (A.8)$$

$$[Z_{14}] = \frac{\eta_0 \omega}{4c} \int_{x_{m1}-\Delta_x/2}^{x_{m1}+\Delta_x/2} \int_{x_{n4}-\Delta_x/2}^{x_{n4}+\Delta_x/2} \int_{y_{n4}-\Delta_y/2}^{y_{n4}+\Delta_y/2} H_0^{(2)}\left(\frac{\omega}{c}\sqrt{(x-x')^2 + (0-y')^2}\right) dy' dx' dx$$

$$[Z_{15}] = \frac{\eta_0 \omega}{4c} \int_{x_{m1}-\Delta_x/2}^{x_{m1}+\Delta_x/2} \int_{x_{n5}-\Delta_x/2}^{x_{n5}+\Delta_x/2} H_0^{(2)}\left(\frac{\omega}{c}\sqrt{(x-x')^2 + (d_2+d_4)^2}\right) dx' dx$$

$$[Z_{21}] = \frac{\eta_0 \omega}{4c} \int_{x_{m2}-\Delta_x/2}^{x_{m2}-\Delta_x/2} \int_{x_{m2}-\Delta_y/2}^{x_{m2}+\Delta_y/2} \int_{x_{n1}-\Delta_x/2}^{x_{n1}+\Delta_x/2} H_0^{(2)}\left(\frac{\omega}{c}\sqrt{(x-x')^2+(y-y')^2}\right) dx'dydx$$

$$[Z_{22}] = \frac{\eta_0 \omega}{4c} \int_{x_{m2}-\Delta_x/2}^{x_{m2}-\Delta_x/2} \int_{x_{m2}-\Delta_y/2}^{x_{m2}+\Delta_y/2} \int_{x_{n2}-\Delta_x/2}^{x_{n2}+\Delta_x/2} \int_{y_{n2}-\Delta_y/2}^{y_{n2}+\Delta_y/2} H_0^{(2)}\left(\frac{\omega}{c}\sqrt{(x-x')^2+(y-y')^2}\right) dy'dx'dydx$$

$$[Z_{23}] = \frac{\eta_0 \omega}{4c} \int_{x_{m2}-\Delta_x/2}^{x_{m2}-\Delta_x/2} \int_{x_{m2}-\Delta_y/2}^{x_{m2}+\Delta_y/2} \int_{x_{n3}-\Delta_x/2}^{x_{n3}+\Delta_x/2} H_0^{(2)}\left(\frac{\omega}{c}\sqrt{(x-x')^2+(y-y')^2}\right) dx'dydx \qquad (A.12)$$

$$[Z_{24}] = \frac{\eta_0 \omega}{4c} \int_{x_{m2}-\Delta_x/2}^{x_{m2}-\Delta_x/2} \int_{x_{m2}-\Delta_y/2}^{x_{m2}+\Delta_y/2} \int_{x_{n4}-\Delta_x/2}^{x_{n4}+\Delta_x/2} \int_{y_{n4}-\Delta_y/2}^{y_{n4}+\Delta_y/2} H_0^{(2)}\left(\frac{\omega}{c}\sqrt{(x-x')^2+(y-y')^2}\right) dy'dx'dydx$$

$$[Z_{25}] = \frac{\eta_0 \omega}{4c} \int_{x_{m2}-\Delta_x/2}^{x_{m2}-\Delta_x/2} \int_{x_{m2}-\Delta_y/2}^{x_{m2}+\Delta_y/2} \int_{x_{n5}-\Delta_x/2}^{x_{n5}+\Delta_x/2} H_0^{(2)}\left(\frac{\omega}{c}\sqrt{(x-x')^2+(y-y')^2}\right) dx'dydx$$

(A.10) and using a Galerkin testing procedure, the integral equation in (A.10) can be discretized and written in matrix form as

$$[V_2]-[Z_{21}][I_1]-[Z_{22}][I_2]-[Z_{23}][I_3] \\ -[Z_{24}][I_4]-[Z_{25}][I_5]=[\eta_2][I_2] \qquad (A.11)$$

where in this case the matrices $[Z_{ij}]$ are given in (A.12). The self-impedance term occurs whenever $i=j$ and $m=n$ simultaneously in (A.12). For this case, the impedance matrix element is calculated using Adaptive Gaussian Quadrature numerical integrations routines as closed form expressions are not available. The integral equation for the fourth layer is similar to (A.10).

Constructing the integral equations for the remaining layers produces five total coupled integral equations. Discretization of each of the integral equations generates the matrix equations below.

$$\sum_{i=1}^{5}\sum_{j=1}^{5}\left([V_i]-[Z_{ij}][I_j]=[\eta_i][I_i]\right) \qquad (A.13)$$

where $i=1,2,3,4,5$. The matrices $[\eta_i]$ are diagonal matrices. Note, since the ground plane on layer 5 is perfectly conducting, the surface impedance $[\eta_5]$ can be set to zero. Equation (A.13) can be written in block matrix form as

$$\begin{bmatrix}[V_1]\\[V_2]\\[V_3]\\[V_4]\\[V_5]\end{bmatrix}=\begin{bmatrix}[Z_{11}]+[\eta_1] & [Z_{12}] & [Z_{13}] & [Z_{14}] & [Z_{15}]\\ [Z_{21}] & [Z_{22}]+[\eta_2] & [Z_{23}] & [Z_{24}] & [Z_{25}]\\ [Z_{31}] & [Z_{32}] & [Z_{33}]+[\eta_3] & [Z_{34}] & [Z_{35}]\\ [Z_{41}] & [Z_{42}] & [Z_{43}] & [Z_{44}]+[\eta_4] & [Z_{45}]\\ [Z_{51}] & [Z_{52}] & [Z_{53}] & [Z_{54}] & [Z_{55}]\end{bmatrix}\begin{bmatrix}[I_1]\\[I_2]\\[I_3]\\[I_4]\\[I_5]\end{bmatrix}$$
(A.14)

The voltage vectors appearing in (A.14) are found either from

$$[V_i] = \int_{x_{m_i}-\Delta_x/2}^{x_{m_i}+\Delta_x/2} E_i^{inc}(x,y)dx \qquad (A.15)$$

where $y$ is equal to $0$, $-d_2$, or $-(d_2+d_4)$ when $i$ is 1, 3, or 5, respectively, or from

$$[V_i] = \int_{x_{m_i}-\Delta_x/2}^{x_{m_i}+\Delta_x/2}\int_{y_{m_i}+\Delta_x/2}^{y_{m_i}+\Delta_x/2} E_i^{inc}(x,y)dydx \qquad (A.16)$$

where $-d_2<y<0$ or $-d_4<y<-d_2$ when $i$ is 2 or 4, respectively. The vectors $I_i$ are vectors of dimension $N_i \times 1$ and represent the unknown coefficients of the expansions given in (A.3) and (A.5).

The desired total aperture electric field defined in (3) is converted to matrix form by

$$[W_T] = \int_{x_{m_1}-\Delta_x/2}^{x_{m_1}+\Delta_x/2} E_d^{tot}(\vec{\rho}_T,\omega)dx \qquad (A.17)$$

### B. Construction of Numerical Green's Function

The numerical Green's function appearing in (5) and (6) can be constructed from the matrix equation (A.14). In the formulations that follow, a superscript $\omega_a$ indicates quantities calculated at frequency $\omega_a$, and a superscript $\omega_b$ indicates quantities calculated at frequency $\omega_b$. Because the numerical Green's function determines the electric field on the top metasurface (layer 1) or the bottom metasurface (layer 3) due to a known source on the top metasurface or the bottom metasurface radiating in the presence of the grounded dielectric spacers only, we may write (A.14) as

$$\begin{bmatrix}[V_1^{\omega_{a/b}}]\\[V_2^{\omega_{a/b}}]\\[V_3^{\omega_{a/b}}]\\[V_4^{\omega_{a/b}}]\\[V_5^{\omega_{a/b}}]\end{bmatrix}=\begin{bmatrix}[Z_{11}^{\omega_{a/b}}] & [Z_{12}^{\omega_{a/b}}] & [Z_{13}^{\omega_{a/b}}] & [Z_{14}^{\omega_{a/b}}] & [Z_{15}^{\omega_{a/b}}]\\ [Z_{21}^{\omega_{a/b}}] & [Z_{22}^{\omega_{a/b}}]' & [Z_{23}^{\omega_{a/b}}] & [Z_{24}^{\omega_{a/b}}] & [Z_{25}^{\omega_{a/b}}]\\ [Z_{31}^{\omega_{a/b}}] & [Z_{32}^{\omega_{a/b}}] & [Z_{33}^{\omega_{a/b}}] & [Z_{34}^{\omega_{a/b}}] & [Z_{35}^{\omega_{a/b}}]\\ [Z_{41}^{\omega_{a/b}}] & [Z_{42}^{\omega_{a/b}}] & [Z_{43}^{\omega_{a/b}}] & [Z_{44}^{\omega_{a/b}}]' & [Z_{45}^{\omega_{a/b}}]\\ [Z_{51}^{\omega_{a/b}}] & [Z_{52}^{\omega_{a/b}}] & [Z_{53}^{\omega_{a/b}}] & [Z_{54}^{\omega_{a/b}}] & [Z_{55}^{\omega_{a/b}}]\end{bmatrix}\begin{bmatrix}[I_1^{\omega_{a/b}}]\\[I_2^{\omega_{a/b}}]\\[I_3^{\omega_{a/b}}]\\[I_4^{\omega_{a/b}}]\\[I_5^{\omega_{a/b}}]\end{bmatrix}$$
(A.18)

where

$$[Z_{22}^{\omega_{a/b}}]' = [Z_{22}^{\omega_{a/b}}]+[\eta_2^{\omega_{a/b}}] \\ [Z_{44}^{\omega_{a/b}}]' = [Z_{44}^{\omega_{a/b}}]+[\eta_4^{\omega_{a/b}}] \qquad (A.19)$$

with

$$[\eta_2^{\omega_{a/b}}] = \frac{1}{j\omega_{a/b}\varepsilon_0(1-\varepsilon_{r2})} \\ [\eta_4^{\omega_{a/b}}] = \frac{1}{j\omega_{a/b}\varepsilon_0(1-\varepsilon_{r4})} \qquad (A.20)$$

To find $E(\vec{\rho}_T, J_T, \omega_{a/b})$, we need to find the field on the top metasurface due to a known $J_T$. Thus, we need to solve

$$\begin{bmatrix} [Z_{11}^{\omega_{a/b}}] & [Z_{12}^{\omega_{a/b}}] & [Z_{13}^{\omega_{a/b}}] & [Z_{14}^{\omega_{a/b}}] & [Z_{15}^{\omega_{a/b}}] \\ [Z_{21}^{\omega_{a/b}}] & [Z_{22}^{\omega_{a/b}}]' & [Z_{23}^{\omega_{a/b}}] & [Z_{24}^{\omega_{a/b}}] & [Z_{25}^{\omega_{a/b}}] \\ [Z_{31}^{\omega_{a/b}}] & [Z_{32}^{\omega_{a/b}}] & [Z_{33}^{\omega_{a/b}}] & [Z_{34}^{\omega_{a/b}}] & [Z_{35}^{\omega_{a/b}}] \\ [Z_{41}^{\omega_{a/b}}] & [Z_{42}^{\omega_{a/b}}] & [Z_{43}^{\omega_{a/b}}] & [Z_{44}^{\omega_{a/b}}]' & [Z_{45}^{\omega_{a/b}}] \\ [Z_{51}^{\omega_{a/b}}] & [Z_{52}^{\omega_{a/b}}] & [Z_{53}^{\omega_{a/b}}] & [Z_{54}^{\omega_{a/b}}] & [Z_{55}^{\omega_{a/b}}] \end{bmatrix} \begin{bmatrix} [I_1^{\omega_{a/b}}] \\ [I_2^{\omega_{a/b}}] \\ [0] \\ [I_4^{\omega_{a/b}}] \\ [I_5^{\omega_{a/b}}] \end{bmatrix} = \begin{bmatrix} [0] \\ [0] \\ [0] \\ [0] \\ [0] \end{bmatrix}$$
(A.21)

or

$$\begin{bmatrix} [Z_{22}^{\omega_{a/b}}]' & [Z_{24}^{\omega_{a/b}}] & [Z_{25}^{\omega_{a/b}}] \\ [Z_{42}^{\omega_{a/b}}] & [Z_{44}^{\omega_{a/b}}]' & [Z_{45}^{\omega_{a/b}}] \\ [Z_{52}^{\omega_{a/b}}] & [Z_{54}^{\omega_{a/b}}] & [Z_{55}^{\omega_{a/b}}] \end{bmatrix} \begin{bmatrix} [I_2^{\omega_{a/b}}] \\ [I_4^{\omega_{a/b}}] \\ [I_5^{\omega_{a/b}}] \end{bmatrix} = \begin{bmatrix} -[Z_{21}^{\omega_{a/b}}][I_1^{\omega_{a/b}}] \\ -[Z_{41}^{\omega_{a/b}}][I_1^{\omega_{a/b}}] \\ -[Z_{51}^{\omega_{a/b}}][I_1^{\omega_{a/b}}] \end{bmatrix}$$

$$= \begin{bmatrix} -[Z_{21}^{\omega_{a/b}}] \\ -[Z_{41}^{\omega_{a/b}}] \\ -[Z_{51}^{\omega_{a/b}}] \end{bmatrix} [I_1^{\omega_{a/b}}]$$
(A.22)

Solve for $[I_2^{\omega_{a/b}}]$, $[I_4^{\omega_{a/b}}]$, and $[I_5^{\omega_{a/b}}]$ due to $[I_1^{\omega_{a/b}}]$

$$\begin{bmatrix} [I_2^{\omega_{a/b}}] \\ [I_4^{\omega_{a/b}}] \\ [I_5^{\omega_{a/b}}] \end{bmatrix} = \begin{bmatrix} [Z_{22}^{\omega_{a/b}}]' & [Z_{24}^{\omega_{a/b}}] & [Z_{25}^{\omega_{a/b}}] \\ [Z_{42}^{\omega_{a/b}}] & [Z_{44}^{\omega_{a/b}}]' & [Z_{45}^{\omega_{a/b}}] \\ [Z_{52}^{\omega_{a/b}}] & [Z_{54}^{\omega_{a/b}}] & [Z_{55}^{\omega_{a/b}}] \end{bmatrix}^{-1} \begin{bmatrix} -[Z_{21}^{\omega_{a/b}}] \\ -[Z_{41}^{\omega_{a/b}}] \\ -[Z_{51}^{\omega_{a/b}}] \end{bmatrix} [I_1^{\omega_{a/b}}]$$
(A.23)

Now we can find $E(\vec{\rho}_T, J_T, \omega_{a/b})$

$$E(\vec{\rho}_T, J_T, \omega_{a/b}) \Rightarrow \begin{bmatrix} [Z_{12}^{\omega_{a/b}}] & [Z_{14}^{\omega_{a/b}}] & [Z_{15}^{\omega_{a/b}}] \end{bmatrix} \begin{bmatrix} [I_2^{\omega_{a/b}}] \\ [I_4^{\omega_{a/b}}] \\ [I_5^{\omega_{a/b}}] \end{bmatrix} + [Z_{11}^{\omega_{a/b}}][I_1^{\omega_{a/b}}]$$

$$= \left( \begin{bmatrix} [Z_{12}^{\omega_{a/b}}] & [Z_{14}^{\omega_{a/b}}] & [Z_{15}^{\omega_{a/b}}] \end{bmatrix} \begin{bmatrix} [Z_{22}^{\omega_{a/b}}]' & [Z_{24}^{\omega_{a/b}}] & [Z_{25}^{\omega_{a/b}}] \\ [Z_{42}^{\omega_{a/b}}] & [Z_{44}^{\omega_{a/b}}]' & [Z_{45}^{\omega_{a/b}}] \\ [Z_{52}^{\omega_{a/b}}] & [Z_{54}^{\omega_{a/b}}] & [Z_{55}^{\omega_{a/b}}] \end{bmatrix}^{-1} \begin{bmatrix} -[Z_{21}^{\omega_{a/b}}] \\ -[Z_{41}^{\omega_{a/b}}] \\ -[Z_{51}^{\omega_{a/b}}] \end{bmatrix} + [Z_{11}^{\omega_{a/b}}] \right) [I_1^{\omega_{a/b}}]$$

$$= [G_{TT}][I_1^{\omega_{a/b}}]$$
(A.24)

The electric field on the bottom metasurface due to $J_T$ can also be found as

$$E(\vec{\rho}_B, J_T, \omega_{a/b}) \Rightarrow \begin{bmatrix} [Z_{32}^{\omega_{a/b}}] & [Z_{34}^{\omega_{a/b}}] & [Z_{35}^{\omega_{a/b}}] \end{bmatrix} \begin{bmatrix} [I_2^{\omega_{a/b}}] \\ [I_4^{\omega_{a/b}}] \\ [I_5^{\omega_{a/b}}] \end{bmatrix} + [Z_{31}^{\omega_{a/b}}][I_1^{\omega_{a/b}}]$$

$$= \left( \begin{bmatrix} [Z_{32}^{\omega_{a/b}}] & [Z_{34}^{\omega_{a/b}}] & [Z_{35}^{\omega_{a/b}}] \end{bmatrix} \begin{bmatrix} [Z_{22}^{\omega_{a/b}}]' & [Z_{24}^{\omega_{a/b}}] & [Z_{25}^{\omega_{a/b}}] \\ [Z_{42}^{\omega_{a/b}}] & [Z_{44}^{\omega_{a/b}}]' & [Z_{45}^{\omega_{a/b}}] \\ [Z_{52}^{\omega_{a/b}}] & [Z_{54}^{\omega_{a/b}}] & [Z_{55}^{\omega_{a/b}}] \end{bmatrix}^{-1} \begin{bmatrix} -[Z_{21}^{\omega_{a/b}}] \\ -[Z_{41}^{\omega_{a/b}}] \\ -[Z_{51}^{\omega_{a/b}}] \end{bmatrix} + [Z_{31}^{\omega_{a/b}}] \right) [I_1^{\omega_{a/b}}]$$

$$= [G_{BT}][I_1^{\omega_{a/b}}]$$
(A.25)

Similarly, one can find fields on the top metasurface (layer 1) and the bottom metasurface (layer 3) due to a known $J_B$. This gives rise to definitions for $[G_{TB}]$ and $[G_{BB}]$.

We also need to find $E^{inc}(\vec{\rho}_T, \omega_{a/b})$, the incident field in the presence of the structure without metallization by solving

$$\begin{bmatrix} [Z_{22}^{\omega_{a/b}}]' & [Z_{24}^{\omega_{a/b}}] & [Z_{25}^{\omega_{a/b}}] \\ [Z_{42}^{\omega_{a/b}}] & [Z_{44}^{\omega_{a/b}}]' & [Z_{45}^{\omega_{a/b}}] \\ [Z_{52}^{\omega_{a/b}}] & [Z_{54}^{\omega_{a/b}}] & [Z_{55}^{\omega_{a/b}}] \end{bmatrix} \begin{bmatrix} [I_2^{\omega_{a/b}}] \\ [I_4^{\omega_{a/b}}] \\ [I_5^{\omega_{a/b}}] \end{bmatrix} = \begin{bmatrix} [V_2^{\omega_{a/b}}] \\ [V_4^{\omega_{a/b}}] \\ [V_5^{\omega_{a/b}}] \end{bmatrix}$$

$$\begin{bmatrix} [I_2^{\omega_{a/b}}] \\ [I_4^{\omega_{a/b}}] \\ [I_5^{\omega_{a/b}}] \end{bmatrix} = \begin{bmatrix} [Z_{22}^{\omega_{a/b}}]' & [Z_{24}^{\omega_{a/b}}] & [Z_{25}^{\omega_{a/b}}] \\ [Z_{42}^{\omega_{a/b}}] & [Z_{44}^{\omega_{a/b}}]' & [Z_{45}^{\omega_{a/b}}] \\ [Z_{52}^{\omega_{a/b}}] & [Z_{54}^{\omega_{a/b}}] & [Z_{55}^{\omega_{a/b}}] \end{bmatrix}^{-1} \begin{bmatrix} [V_2^{\omega_{a/b}}] \\ [V_4^{\omega_{a/b}}] \\ [V_5^{\omega_{a/b}}] \end{bmatrix}$$
(A.26)

for the induced currents in the dielectric and ground plane layers. Then the incident field can then be found

$$[V_T^{\omega_{a/b}}] = \begin{bmatrix} [Z_{32}^{\omega_{a/b}}] & [Z_{34}^{\omega_{a/b}}] & [Z_{35}^{\omega_{a/b}}] \end{bmatrix} \begin{bmatrix} [I_2^{\omega_{a/b}}] \\ [I_4^{\omega_{a/b}}] \\ [I_5^{\omega_{a/b}}] \end{bmatrix} + [V_1^{\omega_{a/b}}]$$

$$= \begin{bmatrix} [Z_{32}^{\omega_{a/b}}] & [Z_{34}^{\omega_{a/b}}] & [Z_{35}^{\omega_{a/b}}] \end{bmatrix} \begin{bmatrix} [Z_{22}^{\omega_{a/b}}]' & [Z_{24}^{\omega_{a/b}}] & [Z_{25}^{\omega_{a/b}}] \\ [Z_{42}^{\omega_{a/b}}] & [Z_{44}^{\omega_{a/b}}]' & [Z_{45}^{\omega_{a/b}}] \\ [Z_{52}^{\omega_{a/b}}] & [Z_{54}^{\omega_{a/b}}] & [Z_{55}^{\omega_{a/b}}] \end{bmatrix}^{-1} \begin{bmatrix} [V_2^{\omega_{a/b}}] \\ [V_4^{\omega_{a/b}}] \\ [V_5^{\omega_{a/b}}] \end{bmatrix} + [V_1^{\omega_{a/b}}]$$
(A.27)

Similarly, one can obtain an expression for $[V_B^{\omega_{a/b}}]$.

ACKNOWLEDGMENT

This work was supported in part by the Office of Naval Research under grant no. N00014-18-1-2536 and the Army Research Office under grant no. W911NF-19-1-0359.

REFERENCES

[1] C. L. Holloway, E. F. Kuester, J. A. Gordon, J. O'Hara, J. Booth and D. R. Smith, "An Overview of the Theory and Applications of Metasurfaces: The Two-Dimensional Equivalents of Metamaterials," *in IEEE Antennas Propag. Mag.*, vol. 54, no. 2, pp. 10-35, April 2012.

[2] E. F. Kuester, M. A. Mohamed, M. Piket-May and C. L. Holloway, "Averaged transition conditions for electromagnetic fields at a metafilm," *in IEEE Antennas Propag. Mag.*, vol. 51, no. 10, pp. 2641-2651, Oct. 2003.

[3] J. Budhu and A. Grbic, "Recent Advances in Bianisotropic Boundary Conditions: Theory, Capabilities, Realizations, and Applications," *arXiv:2108.05965v2 [physics.app-ph]*.

[4] C. Pfeiffer, C. Zhang, V. Ray, L. J. Guo, A. Grbic, "Polarization rotation with ultra-thin bianisotropic metasurfaces," *Optica*, vol. 3, no. 4, pp. 427-432, April 2016.

[5] C. Pfeiffer and A. Grbic, "Millimeter-Wave Transmitarrays for Wavefront and Polarization Control," *in IEEE Trans. Microw. Theory Tech.*, vol. 61, no. 12, pp. 4407-4417, Dec. 2013.

[6] C. Pfeiffer and A. Grbic "Bianisotropic metasurfaces for optimal polarization control: Analysis and synthesis," *Physical Review Applied*, vol. 2, 044011, October 2014.

[7] J. Budhu and A. Grbic, "A Rigorous Approach to Designing Reflectarrays," *2019 23rd International Conference on Applied Electromagnetics and Communications (ICECOM)*, 2019, pp. 1-3.

[8] J. Budhu and A. Grbic, "Passive Reflective Metasurfaces for Far-Field Beamforming," *2021 15th European Conference on Antennas and Propagation (EuCAP)*, 2021, pp. 1-4.

[9] V. S. Asadchy, M. Albooyeh, S. N. Tcvetkova, A. Díaz-Rubio, Y. Ra'di, and S. A. Tretyakov, "Perfect control of reflection and refraction using spatially dispersive metasurfaces," *Phys. Rev. B* 94, 075142.

[10] A. Díaz-Rubio*, V. S. Asadchy, A. Elsakka and S.A. Tretyakov, "From the generalized reflection law to the realization of perfect anomalous reflectors," *Science Advances* 11 Aug 2017:Vol. 3, no. 8, e1602714.

[11] A. Epstein and G. V. Eleftheriades, "Synthesis of Passive Lossless Metasurfaces Using Auxiliary Fields for Reflectionless Beam Splitting and Perfect Reflection," *Phys. Rev. Lett.*, vol. 117, no. 25, p. 256103, Dec. 2016

[12] J. Budhu and A. Grbic, "Perfectly Reflecting Metasurface Reflectarrays: Mutual Coupling Modeling Between Unique Elements Through Homogenization," *in IEEE Trans. Antennas Propag.*, vol. 69, no. 1, pp. 122-134, Jan. 2021.

[13] J. Budhu and A. Grbic, "A Reflective Metasurface for Perfect Cylindrical to Planar Wavefront Transformation," *2020 Fourteenth International Congress on Artificial Materials for Novel Wave Phenomena (Metamaterials)*, 2020, pp. 234-236.

[14] M. Del Mastro, M. Ettorre and A. Grbic, "Dual-Band, Orthogonally-Polarized LP-to-CP Converter for SatCom Applications," *in IEEE Trans. Antennas Propag.*, vol. 68, no. 9, pp. 6764-6776, Sept. 2020.

[15] N. J. G. Fonseca and C. Mangenot, "Low-profile polarizing surface with dual-band operation in orthogonal polarizations for broadband satellite applications," *The 8th European Conference on Antennas and Propagation (EuCAP 2014)*, 2014, pp. 471-475.

[16] Khan, B., Kamal, B., Ullah, S. et al., "Design and experimental analysis of dual-band polarization converting metasurface for microwave applications." *Sci Rep* 10, 15393 (2020).

[17] Yizhuo Yu, Fajun Xiao, Chong He, Ronghong Jin, and Weiren Zhu, "Double-arrow metasurface for dual-band and dual-mode polarization conversion," *Opt. Express* 28, 11797-11805 (2020).

[18] L. Huang, et. al., "Bilayer Metasurfaces for Dual- and Broadband Optical Antireflection," *ACS Photonics* 2017, 4, 9, 2111–2116.

[19] P. Naseri and S. V. Hum, "A Generative Machine Learning-Based Approach for Inverse Design of Multilayer Metasurfaces," *arXiv:2008.02074 [eess.SP]*.

[20] W. Guo, G. Wang, H. Hou, K. Chen, and Y. Feng, "Multi-functional coding metasurface for dual-band independent electromagnetic wave control," *Opt. Express* 27, 19196-19211 (2019).

[21] P. Vinod Kumar and B. Ghosh, "A Dual-Band Multi-layer Metasurface Lens," *2018 IEEE Indian Conference on Antennas and Propogation (InCAP)*, Hyderabad, India, 2018, pp. 1-4.

[22] C. Han, J. Huang and Kai Chang, "A high efficiency offset-fed X/ka-dual-band reflectarray using thin membranes," *in IEEE Trans. Antennas Propag.*, vol. 53, no. 9, pp. 2792-2798, Sept. 2005.

[23] C. Han, C. Rodenbeck, J. Huang and Kai Chang, "A C/ka dual frequency dual Layer circularly polarized reflectarray antenna with microstrip ring elements," *in IEEE Trans. Antennas Propag.*, vol. 52, no. 11, pp. 2871-2876, Nov. 2004.

[24] D. I. Wu, R. C. Hall and J. Huang, "Dual-frequency microstrip reflectarray," *IEEE Antennas and Propagation Society International Symposium*. 1995 Digest, 1995, pp. 2128-2131 vol.4.

[25] R. Xie, Y. Liu, T. Wang, G. Zhai, J. Gao and J. Ding, "High-efficiency Dual-band Bifocal Metalens Based on Reflective Metasurface," *2019 IEEE International Conference on Computational Electromagnetics (ICCEM)*, 2019, pp. 1-3.

[26] D. R. Prado, M. Arrebola, M. R. Pino and G. Goussetis, "Contoured-Beam Dual-Band Dual-Linear Polarized Reflectarray Design Using a Multiobjective Multistage Optimization," *in IEEE Trans. Antennas Propag.*, vol. 68, no. 11, pp. 7682-7687, Nov. 2020.

[27] G. Xu, S. V. Hum and G. V. Eleftheriades, "Dual-Band Reflective Metagratings With Interleaved Meta-Wires," *in IEEE Trans. Antennas Propag.*, vol. 69, no. 4, pp. 2181-2193, April 2021.

[28] B. Xi, Q. Xue, Y. Cai, L. Bi, and Y. Wang, "Analytical Method for an FSS-Sandwiched Dual-Band Reflectarray Antenna," *Progress In Electromagnetics Research M*, Vol. 77, 61–71, 2019.

[29] Faenzi, M., Minatti, G., González-Ovejero, D. et al. "Metasurface Antennas: New Models, Applications and Realizations." *Sci Rep* 9, 10178 (2019).

[30] W. Gibson, <u>The Method of Moments in Electromagnetics</u>, Chapman & Hall/CRC, Taylor and Francis Group, Boca Raton, FL, 2008.

[31] J. Budhu, A. Grbic and E. Michielssen, "Design of Multilayer, Dualband Metasurface Reflectarrays," *2020 14th European Conference on Antennas and Propagation (EuCAP)*, 2020, pp. 1-4.

[32] J. Budhu, A. Grbic and E. Michielssen, "Dualband Stacked Metasurface Reflectarray," *2020 IEEE International Symposium on Antennas and Propagation and North American Radio Science Meeting*, 2020, pp. 821-822.

[33] M. Bodehou, C. Craeye and I. Huynen, "Multifrequency Band Synthesis of Modulated Metasurface Antennas," *in IEEE Antennas and Wireless Propagation Letters*, vol. 19, no. 1, pp. 134-138, Jan. 2020.

[34] A. Epstein and G. V. Eleftheriades, "Arbitrary Power-Conserving Field Transformations With Passive Lossless Omega-Type Bianisotropic Metasurfaces," *in IEEE Trans. Antennas Propag.*, vol. 64, no. 9, pp. 3880-3895, Sept. 2016.